\documentclass[aip,pop,preprint,floatfix,superscriptaddress,amsmath,amsfonts,amssymb]{revtex4-1}
\usepackage{amsfonts}
\usepackage{amsmath}
\usepackage{color,graphicx}
\usepackage[colorlinks,linkcolor=blue,citecolor=blue,bookmarks,pdfstartview=FitH]{hyperref}

\newcommand\myfigure{./}

\begin{document}

\title{Nonlinear dynamics of Shear Alfv\'en fluctuations in Divertor Tokamak Test facility plasmas}

\author{T. Wang}
\affiliation{Institute for Fusion Theory and Simulation and Department of Physics, Zhejiang University, Hangzhou 310027, China}

\author{X. Wang}
\affiliation{Max-Planck-Institut f\"ur Plasmaphysik, Boltzmannstr. 2, 85748 Garching, Germany}

\author{S. Briguglio}
\affiliation{ENEA, Fusion and Nuclear Safety Department, C. R. Frascati, Via E. Fermi 45, 00044 Frascati (Roma), Italy}

\author{Z. Qiu}
\affiliation{Institute for Fusion Theory and Simulation and Department of Physics, Zhejiang University, Hangzhou 310027, China}

\author{G. Vlad}
\affiliation{ENEA, Fusion and Nuclear Safety Department, C. R. Frascati, Via E. Fermi 45, 00044 Frascati (Roma), Italy}

\author{F. Zonca}
\email{fulvio.zonca@enea.it}
\affiliation{ENEA, Fusion and Nuclear Safety Department, C. R. Frascati, Via E. Fermi 45, 00044 Frascati (Roma), Italy}
\affiliation{Institute for Fusion Theory and Simulation and Department of Physics, Zhejiang University, Hangzhou 310027, China}
\begin{abstract}
Following the analysis on linear spectra of shear Alfv\'en fluctuations excited by energetic particles (EPs) in the Divertor Tokamak Test (DTT) facility plasmas [T. Wang {\it et al.}, Phys. Plasmas {\bf 25}, 062509 (2018)], in 
this work, nonlinear dynamics of the corresponding mode saturation and the fluctuation induced EP transport is studied by hybrid magnetohydrodynamic-gyrokinetic simulations.
For the reversed shear Alfv\'en eigenmode driven by magnetically trapped EP precession resonance in the central core region of DTT plasmas, the saturation is mainly due to radial decoupling of resonant trapped EPs.
Consistent with the wave-EP resonance structure, EP transport occurs in a similar scale to the mode width.
On the other hand, passing EP transport is analyzed in detail for toroidal Alfv\'en eigenmode in the outer core region, with mode drive from both passing and trapped EPs.
It is shown that passing EPs experience only weak redistributions in the weakly unstable case; and the transport extends to meso-scale diffusion in the strongly unstable case, due to orbit stochasticity induced by resonance overlap.
Here, weakly/strongly unstable regime is determined by Chirikov condition for resonance overlap.
This work then further illuminates rich and diverse nonlinear EP dynamics related to burning plasma studies, and the capability of DTT to address these key physics.
\end{abstract}

\maketitle

\section{Introduction}
\label{sec:intro}
In Ref.~\onlinecite{paper1}, we have investigated linear dynamics of shear Alfv\'en fluctuations excited by energetic particles (EPs)
in the recently proposed next generation tokamak device, the Divertor Tokamak Test (DTT) facility,\cite{albanese17a}
which mainly aims at studying viable divertor configurations for the demonstration power plant (DEMO).\cite{chapman11}
Despite the practical objectives of DTT,
we have shown that many substantial physics can be explored in DTT core plasmas,\cite{paper1} including fundamental issues related to burning plasma operations.\cite{zonca15a,chen16}
In particular, the EPs produced by nuclear fusion reactions and/or auxiliary heating methods
can drive Alfv\'enic fluctuations unstable via wave-particle resonant interactions, as their characteristic dynamic frequencies are in the magnetohydrodynamic (MHD) range.
Depending on the intensity of EP drive and resonance condition, the collective fluctuations of shear Alfv\'en waves (SAWs) could exist as Alfv\'en eigenmodes (AEs)\cite{cheng85} inside frequency gaps of SAW continuous spectrum (continuum), or as energetic particle continuum modes (EPMs).\cite{chen94}
Following the linear analysis presented in Ref.~\onlinecite{paper1},
in this paper, we focus on the nonlinear saturation of the Alfv\'enic fluctuations and the fluctuation induced EP transport in DTT plasmas,
as the assessment of EP confinement property is crucial in the next generation tokamak experiments.
In fact, only a small fraction of EP loss could be tolerated in burning plasma devices without severely damaging the plasma facing components.
Thus, a deeper comprehension of these key physics is not only important for the success of DTT, but also of practical interest for burning plasma studies, such as in the International Thermonuclear Experimental Reactor (ITER)\cite{IPB99_chap5,fasoli07,pinches15} and DEMO.\cite{chapman11}

Targeting self-sustained nuclear fusion in the next generation tokamaks,
the physics understanding of SAW-EP dynamics in toroidal plasmas has significantly improved in the last several decades,
and is reviewed in a few recent publications\cite{fasoli07,heidbrink08a,breizman11,lauber13,sharapov13,gorelenkov14,chen16} from the perspective of theoretical and experimental research as well as numerical simulation.
In general, the nonlinear saturation of SAW fluctuation may follow two routes, i.e., nonlinear wave-wave and wave-EP interactions.\cite{chen13,chen16}
In this paper, we focus on the latter route by means of hybrid MHD-gyrokinetic code (HMGC)\cite{briguglio95,briguglio98} simulations, due to the crucial role played by EPs in the multi-scale dynamics of fusion plasmas.\cite{zonca15a}
Adopting the theoretical framework of the general fishbone-like dispersion relation,\cite{zonca14a,zonca14b,chen16}
nonlinear dynamics and saturation of a single toroidal mode number SAW fluctuation can be understood as two fundamental mechanisms, namely, resonance detuning and radial decoupling.\cite{zonca13,zonca14c,briguglio14,zonca15b,chen16}
Briefly speaking, resonance detuning is due to the nonlinear shift in the wave-EP phase, and it is ubiquitous in wave-particle resonant interactions.
On the other hand, radial decoupling is due to the finite nonlinear excursion of EP orbits with respect to the likewise finite localized mode structure in nonuniform plasmas.\cite{zonca15b,chen16}
Thus, in a realistic plasma, the complex behavior underlying the nonlinear interplay between SAW fluctuation and EPs depends on the relative importance of the two mechanisms.
As shown theoretically\cite{zonca15b,chen16} and by recent numerical simulations,\cite{zhang12,wang12,vlad13,briguglio14,wang16a,wang16b,vlad16,briguglio17} the saturation mechanism is determined by the relative ordering of nonlinear EP orbit excursion to the perpendicular (with respect to equilibrium magnetic field) fluctuation wavelength and/or equilibrium nonuniformity; and it can be reflected by the relative scale lengths of wave-EP power transfer, mode structure and effective resonance condition.\cite{briguglio14,wang16a,wang16b,briguglio17}
For two paradigmatic cases, typically in the marginally unstable limit, nonlinear EP orbit excursion is restricted by the effective resonance condition, and is much smaller than the perpendicular fluctuation wavelength;
that is, the resonant EP response is similar to that of a uniform plasma.
Hence, in this regime, resonance detuning outweighs radial decoupling and, when only resonance detuning is considered, the saturated fluctuation amplitude scales quadratically with respect to the linear growth rate of the mode,\cite{berk92b,breizman93} consistent with that predicted by wave-particle trapping mechanism typical of a 1-D beam-plasma system.\cite{oneil71}
Meanwhile, in the strongly unstable regime with non-perturbative EP response, the EP orbit excursion is comparable with the perpendicular fluctuation wavelength, radial nonuniformity becomes essential for the resonant EP response, and radial decoupling is therefore crucially important.\cite{zonca15b,chen16}
As shown by previous numerical simulations and predicted by analytic models, the scaling of saturation amplitude versus mode linear growth rate could be linear in this regime.\cite{wang16a,wang16b,briguglio17}
In all cases, the non-perturbative EP response and plasma nonuniformity can introduce additional twists in the complex behaviors underlying wave-EP power exchange, and allow enhanced fluctuation levels with respect to the predicted quadratic and/or linear scaling.\cite{zonca15a,chen16}
Thus, the proper description of saturation mechanism generally requires accounting for the self-consistent interplay of mode structures and EP transport, as extensively discussed in the recent comprehensive review paper by Chen and Zonca.\cite{chen16}

In this work, some of the key integrated physics aspects of burning plasmas are addressed for the DTT reference scenario. DTT plasmas can be generally divided into a central core region, characterized by low magnetic shear and coherent fluctuation induced redistributions of magnetically trapped EPs; and an outer core region, with finite magnetic shear and predominant diffusive losses of passing EPs due to resonance overlap.

Due to the similarity to ITER of DTT dimensionless parameters relative to both supra-thermal and core plasma components, the Alfv\'enic fluctuation spectrum resonantly excited by EPs is characterized by toroidal mode numbers $n\sim O(10)$\cite{paper1} and, therefore, by micro-scales that are of the same order of the meso-scale structures spontaneously formed by drift wave turbulence.
Thus, DTT core plasmas can access operation regimes where complex behavior will mimic those of reactor relevant fusion plasmas, with EPs acting as mediators of cross scale couplings.\cite{zonca15a,chen16}
This work, in particular, will address and illuminate the rich variety of spatiotemporal scales self-consistently generated in DTT plasmas, and resulting from nonlinear interplay of Alfv\'enic fluctuations and EP sources of various strength.

EP transport will be analyzed in phase space, since AEs excited in the central core region by magnetically trapped EPs are characterized by very different resonance structures and corresponding spatiotemporal scales than AEs due to both trapped and passing EPs in the outer core region.\cite{paper1} The different behaviors will be discussed by means of test particle Hamiltonian mapping techniques\cite{vlad13,briguglio14,briguglio17} to illuminate the nonlinear evolution of phase space zonal structures\cite{zonca15b,chen16} and, ultimately, its impact on EP transport, characterized by both coherent nonlinear redistributions as well as diffusive radial fluxes.

This paper is organized as follows.
The simulation model and main simulation parameters are presented in Sec.~\ref{sec:numerical}.
In Sec.~\ref{sec:nonlinear}, we analyze the nonlinear dynamics of two types of SAW fluctuations interacting resonantly with EPs; namely, reversed shear Alfv\'en eigenmodes (RSAEs) resonantly excited by trapped EPs in the central core region, and toroidal Alfv\'en eigenmodes (TAEs) destabilized by both trapped as well as passing EPs in the DTT outer core.
In this work, we focus on the initial mode saturation in order to illustrate and discuss the underlying physics, and to illuminate the richness of diverse nonlinear behaviors that can be expected in the DTT reference scenario.
More complicated long time scale nonlinear evolutions are intentionally left to further and more detailed analyses to be carried out as future work.
Section~\ref{sec:sumdis} gives the final summary and discussion.

\section{Numerical model and simulation parameters}
\label{sec:numerical}
In this paper, we recall the numerical model of the DTT reference scenario considered in Ref.~\onlinecite{paper1}.
Here, we only summarize the most important numerical aspects, while a full description and discussion of the adopted model are given in Ref.~\onlinecite{paper1}.

The simulation code HMGC\cite{briguglio95,briguglio98} is based on the hybrid MHD-gyrokinetic model,\cite{park92}
and focuses on the self-consistent interplay between thermal plasma components and fast/energetic particles in simplified tokamak geometry.
In this study, the bulk plasma fluctuations are described by a set of $O(\epsilon^{3})$-reduced MHD equations\cite{izzo83} in the limit of zero pressure, where $\epsilon \equiv a/R_{0}$ is the inverse aspect ratio, with $a$ and $R_{0}$ the minor and major tokamak radii, respectively.
The EP response is accounted for by solving nonlinear Vlasov equation in the drift-kinetic limit with particle-in-cell method, and enters in the MHD equations non-perturbatively via the pressure coupling formulation.\cite{park92}
Therefore, finite Larmor radius effect is neglected, but finite magnetic drift orbit width effect is fully taken into account.\cite{zonca06a,wang11,zonca14a,zonca14b}
Consistent with Ref.~\onlinecite{paper1}, in this work, we investigate a single toroidal mode number $n$ in each simulation case, while multi-$n$ simulations will be part of future work (see Ref.~\onlinecite{vlad18} for a recent publication on this subject).
Thus, MHD nonlinear mode-mode coupling is neglected, but EP nonlinearities are self-consistently retained.
We also emphasize, as anticipated in the Introduction, that single-$n$ simulations do not necessarily imply studying the nonlinear dynamics of isolated resonances.
Quite the contrary, we will be able to discuss various features of both isolated (Sec.~\ref{subsec:central}) as well as overlapped (Sec.~\ref{subsec:outer}) resonances.

\begin{figure}
\includegraphics{\myfigure 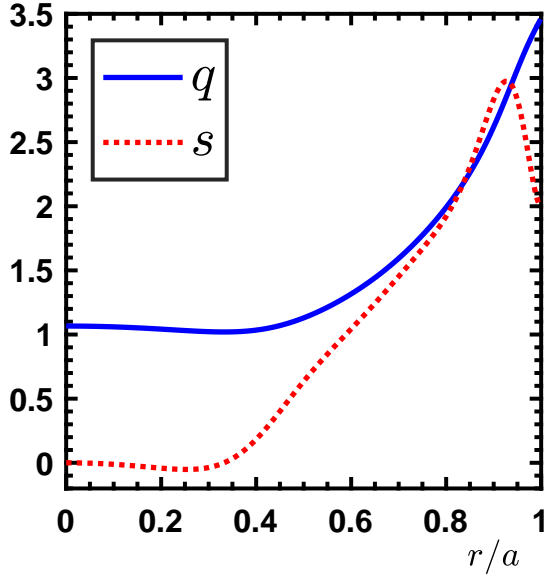}
\caption{Radial profiles of equilibrium safety factor $q$ and magnetic shear $s$.}
\label{fig:equilibrium}
\end{figure}

\begin{figure}
\includegraphics{\myfigure 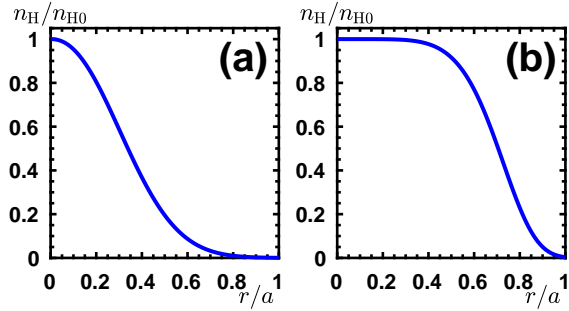}
\caption{Radial profiles of normalized EP density $n_{\rm H}$. The two profiles are used for, respectively, (a) the central core region, and (b) the outer core region.}
\label{fig:EP_profiles}
\end{figure}

A shifted circular equilibrium with $\epsilon=0.18$ is adopted in this paper, along with ITER-like EP parameters.\cite{paper1,pizzuto10}
Figure~\ref{fig:equilibrium}
shows the radial profiles of equilibrium safety factor $q$ and the corresponding magnetic shear $s \equiv rq'/q$, where ``prime'' indicates derivative with respect to the minor radius coordinate $r$.
As articulated in Ref.~\onlinecite{paper1}, the structure of the adopted equilibrium suggests a subdivision into a central core region with $q$ marginally above unity and vanishing $s$, and an outer core region with larger $q$ and finite $s$.
The two regions are investigated separately in this paper, adopting EPs with different radial pressure profiles (cf. Fig.~\ref{fig:EP_profiles}).
EPs are assumed to be fusion born alpha particles characterized by an isotropic slowing-down distribution function.
The pressure drive of EPs is controlled by the normalized radial profile of EP density $n_{\rm H}/n_{\rm H0}$, shown in Fig.~\ref{fig:EP_profiles}
for, respectively, the central and outer core region.
Here, $n_{\rm H0}$ indicates the value of $n_{\rm H}$ on the magnetic axis, and is normalized to on-axis bulk ion density $n_{\rm i0}$ to control the intensity of EP drive (cf. Ref.~\onlinecite{paper1}).
Other important parameters are $\rho_{\rm H}/a=0.01$, $v_{\rm H}/v_{\rm A0}=1.80$,
with $\rho_{\rm H}\equiv v_{\rm H}/\Omega_{\rm H}$ the EP Larmor radius, $v_{\rm H}\equiv\sqrt{E_{0}/m_{\rm H}}$ the characteristic EP birth speed, $\Omega_{\rm H}$ the EP cyclotron frequency, $E_{0}$ the alpha particle birth energy, $m_{\rm H}$ the EP mass, and $v_{\rm A0}$ the on-axis Alfv\'en speed.
Note that $\rho_{\rm H}/a=0.01$ is significantly smaller than in present day tokamaks;\cite{chen16}
and the smaller EP orbit width in DTT is crucial to determine the linear\cite{paper1} as well as nonlinear dynamics of resonantly excited Alfv\'enic fluctuations.
In particular, the range of excited toroidal mode numbers is such that the peculiar role of EPs in cross-scale coupling becomes evident, as to be shown in Sec.~\ref{sec:nonlinear}.

In this paper, in order to illuminate the nonlinear physics anticipated in the Introduction, we focus on the two representative cases discussed in detail in Ref.~\onlinecite{paper1}, namely, $n=4$ RSAE case in the central core region (Sec.~\ref{subsec:central}), and $n=6$ TAE case in the outer core region (Sec.~\ref{subsec:outer}).
The selected two cases can well represent the peculiar features of each region, while the nonlinear dynamics of other modes with different toroidal mode numbers can be predicted following their linear properties\cite{paper1} and the analysis in this paper.
In all simulations, poloidal harmonic $m$ is retained in the interval $[3,~14]$ for $n=4$, and in the interval $[5,~21]$ for $n=6$.

\section{Nonlinear dynamics}
\label{sec:nonlinear}

In this section, after briefly reviewing the linear spectra reported in Ref.~\onlinecite{paper1}, the nonlinear saturation of SAW fluctuations and the associated EP transport in DTT plasmas are investigated by HMGC simulations.
As indicated above,
we analyze $n=4$ RSAE fluctuations for the central core region in Sec.~\ref{subsec:central}, and $n=6$ TAE fluctuations for the outer core region in Sec.~\ref{subsec:outer}.
Normalized EP pressure profiles with variable intensity assumed as initial conditions in the two cases are shown in Fig.~\ref{fig:EP_profiles}(a) and (b), respectively.
The two cases are characterized by very different wave-EP resonance structures.\cite{paper1}
The dominant destabilization mechanism for the RSAE fluctuations is the precession resonance with magnetically trapped EPs;
meanwhile, the wave-EP power transfer for the TAE fluctuations consists of comparable contributions from the precession resonance with trapped EPs, and several transit harmonic resonances with passing EPs.
Different mechanisms of EP transport, suggested by the wave-EP resonance conditions and relevant spatial scales, are also investigated in detail in this section.

\subsection{RSAE nonlinear dynamics}
\label{subsec:central}

\begin{figure}
\includegraphics{\myfigure 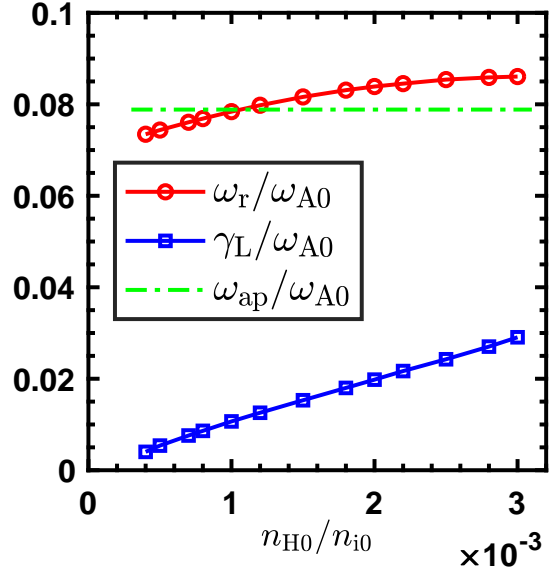}
\caption{RSAE real frequencies $\omega_{\rm r}$ (circles) and linear growth rates $\gamma_{\rm L}$ (squares) shown as functions of EP on-axis density $n_{\rm H0}$.
Here, on-axis Alfv\'en frequency $\omega_{\rm A0}$ is used for normalization.
In addition, the RSAE accumulation point frequency $\omega_{\rm ap}$ is indicated as a horizontal dash-dotted line.}
\label{fig:RSAE_nH}
\end{figure}

\begin{figure}
\includegraphics{\myfigure 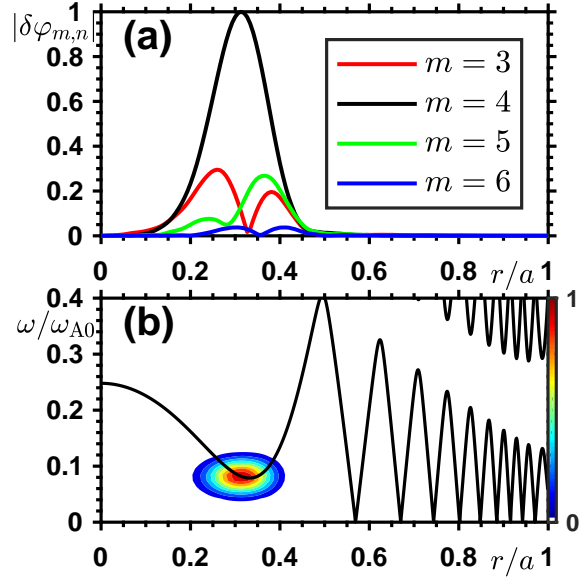}
\caption{Radial structures of several dominant Fourier decomposed poloidal harmonics of scalar potential fluctuation $\delta\varphi_{m,n}$ [frame (a)],
and the intensity contour plot of the corresponding power spectrum in the $(r,\omega)$ plane [frame (b)] shown in arbitrary units
for the RSAE case with $n_{\rm H0}/n_{\rm i0}=1.5\times10^{-3}$ in the linear stage.
The solid curves in frame (b) represent SAW continua.}
\label{fig:RSAE_mode}
\end{figure}

A series of $n=4$ RSAE cases with $n_{\rm H0}/n_{\rm i0}$ in the interval $[0.0004,~0.0030]$ are analyzed.
The unstable fluctuations are characterized by very similar mode structures, but have different spectral properties.
The mode real frequencies $\omega_{\rm r}$ and linear growth rates $\gamma_{\rm L}$ are shown in Fig.~\ref{fig:RSAE_nH},
where the upward frequency shift due to the non-perturbative effect of EPs is evident.\cite{paper1}
Moreover, $\gamma_{\rm L}$ scales almost linearly with $n_{\rm H0}$, suggesting very low instability threshold of EP drive as a result of low background damping.
For a reference case with $n_{\rm H0}/n_{\rm i0}=1.5\times10^{-3}$,
several dominant poloidal harmonics of scalar potential fluctuation $\delta\varphi_{m,n}$ in the linear growth stage are shown in Fig.~\ref{fig:RSAE_mode},
along with the corresponding power spectrum in the $(r,\omega)$ plane.
Consistent with the equilibrium $q$ profile and the structure of SAW continuum, the $m=4$ harmonic is dominant, and the mode is radially localized near the surface with minimum $q$ value.

\begin{figure}
\includegraphics{\myfigure 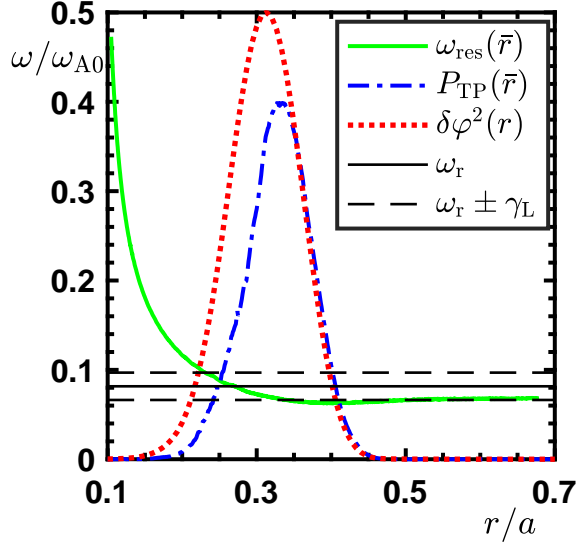}
\caption{For the reference case in the linear stage, the test particle precession resonance frequency $\omega_{\rm res}(\bar r,\ell=0)$ (solid curve) [cf. Eq.~(\ref{eq:omega_res_trapped})] is shown with respect to the orbit averaged radial coordinate $\bar r$.
$\omega_{\rm r}$ and $\omega_{\rm r}\pm\gamma_{\rm L}$ are indicated as, respectively, horizontal solid and dashed lines to illustrate the resonance condition [cf. Eq.~(\ref{eq:resonance})].
In addition, the integrated test particle power transfer $P_{\rm TP}(\bar r)$ (dash-dotted curve), as well as the radial mode structure, denoted by $\delta\varphi^2(r)=\int|\delta\varphi_{m,n}|^2{\rm d}\theta{\rm d}\phi$ (dotted curve), are shown in arbitrary units.}
\label{fig:RSAE_resonance}
\end{figure}

As a useful tool in analyzing wave-EP resonant interactions, test particle method is extensively applied to illustrate the resonance condition as well as the nonlinear dynamics of mode saturation and EP transport.
Test particles are chosen as ``representative'' of resonant EPs, which can be readily identified from wave-EP power transfer.\cite{briguglio14}
Further details about test particle selection are given in Appendix \ref{app:TPs} (interested readers may also refer to Ref.~\onlinecite{briguglio14} for an exhaustive description), while here we only emphasize that the test particle population is characterized by two constants of the perturbed motion, $M$ and $C$, corresponding to a reduced phase space grid (of trapped particles) with significant wave-EP power transfer in the linear stage.
Here, $M$ is the magnetic moment and $C$, given by Eq.~(\ref{eq:C}), is an invariant constructed from the extended phase space Hamiltonian.\cite{white12,zonca15b,chen16}
For the reference case, the linear resonance structure is shown in Fig.~\ref{fig:RSAE_resonance},
along with radial mode structure and flux surface integrated test particle power transfer.
We can see that the effective wave-particle power transfer is limited by the resonance condition on the inner side, and by the radially localized mode structure on the outer side.
The radial width of power transfer (denoted by $\Delta r_{\rm power}$) is similar to that of the mode structure (denoted by $\Delta r_{\rm mode}$);
but it remains much smaller than the effective resonance width $\Delta r_{\rm res}$ [cf. Eq.~(\ref{eq:resonance}) and the discussion below] or equilibrium pressure gradient scale length [cf. Fig.~\ref{fig:EP_profiles}(a)].
Furthermore, similar structures can also be found for the lower growth rate cases reported in Fig.~\ref{fig:RSAE_nH} due to the flat $\omega_{\rm res}$ profile.
Meanwhile, for higher growth rate cases, power transfer is entirely limited by the finite mode width as $\Delta r_{\rm res}$ becomes broader.
In general, $\Delta r_{\rm power} \lesssim \Delta r_{\rm mode} \ll \Delta r_{\rm res}$ for all considered cases,
suggesting that radial decoupling is the dominant mechanism for mode saturation, consistent with the strongly non-perturbative EP response.
In the weakly unstable limit where $\Delta r_{\rm power}<\Delta r_{\rm mode}$, however, resonance detuning and radial decoupling may both play important roles and should be treated on the same footing.

\begin{figure*}
\includegraphics{\myfigure 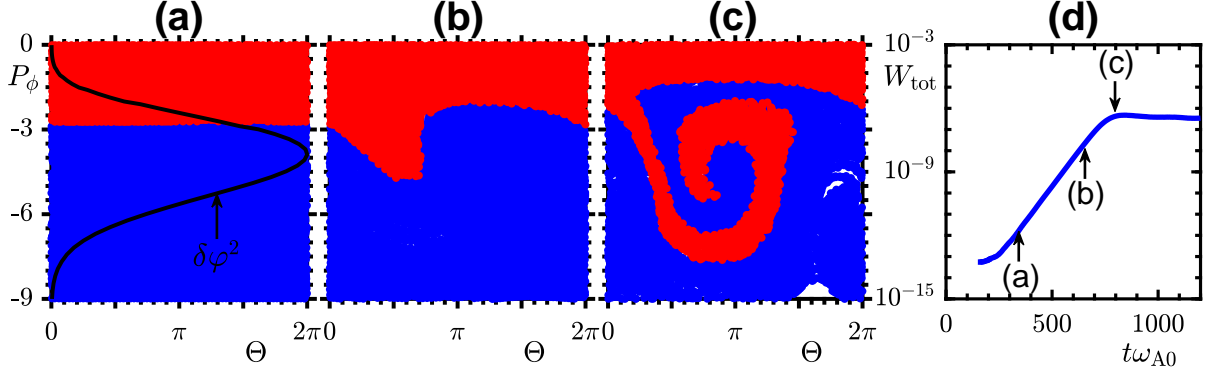}
\caption{Kinetic Poincar\'e plots of test particles shown in the $(\Theta,P_{\phi})$ plane for three successive times of the reference case [frames (a)-(c)]. Frame (d) indicates the three considered times in the time evolution of the total (kinetic plus magnetic) perturbed field energy $W_{\rm tot}$.
In frames (a)-(c), $\Theta$ is modulo $2\pi$, and $P_{\phi}$ is normalized to $m_{\rm H}av_{\rm H}$.
The test particle marker is colored by the particle's initial $P_{\phi}$ value: red for $P_{\phi}>P_{\phi\rm res}$ and blue otherwise.
In addition, radial mode structure $\delta\varphi^2(r)$, given in Fig.~\ref{fig:RSAE_resonance}, is also shown in frame (a) by mapping $\bar r$ into $P_{\phi}$ coordinates.}
\label{fig:RSAE_HM}
\end{figure*}

\begin{figure}
\includegraphics{\myfigure 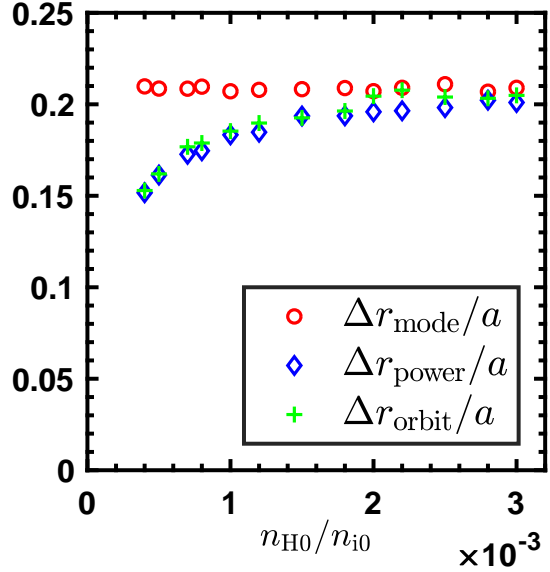}
\caption{For the cases reported in Fig.~\ref{fig:RSAE_nH}, the averaged resonant test particle orbit radial excursion $\Delta r_{\rm orbit}$ at saturation is compared with mode radial width $\Delta r_{\rm mode}$ and power transfer radial width $\Delta r_{\rm power}$ in the linear stage.
Here, $\Delta r_{\rm orbit}$ is calculated as the largest variation of the particle's equatorial plane radial coordinate, and is averaged over the resonant test particle population.
$\Delta r_{\rm mode}$ and $\Delta r_{\rm power}$ are measured as the radial width of the region where the corresponding quantity is larger than the $10\%$ of the peak value (cf. Fig.~\ref{fig:RSAE_resonance}).}
\label{fig:RSAE_delta_r}
\end{figure}

\begin{figure}
\includegraphics{\myfigure 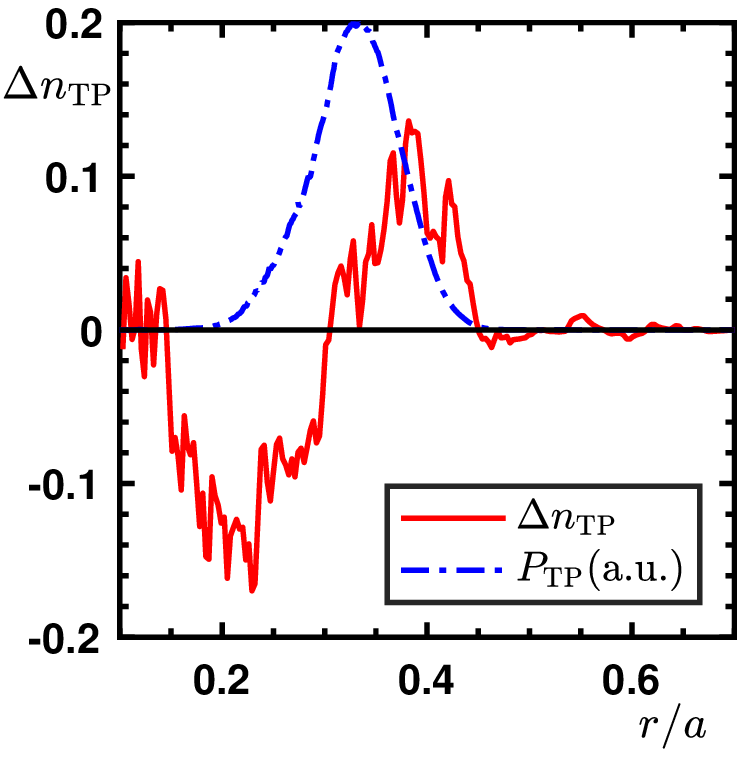}
\caption{The distortion of test particle density radial profile $\Delta n_{\rm TP}$ (solid curve) shown in arbitrary units for the reference case.
$\Delta n_{\rm TP}$ is calculated as the difference of test particle density at saturation with respect to the linear stage, and is normalized to the test particle density at mode peak.
The integrated wave-particle power transfer radial profile $P_{\rm TP}$ (dash-dotted curve) in the linear stage, reported in Fig.~\ref{fig:RSAE_resonance}, is also shown in arbitrary units for comparison.}
\label{fig:RSAE_density}
\end{figure}

The nonlinear evolution of long-lived EP phase space zonal structures can be illustrated by test particle Hamiltonian mapping technique via kinetic Poincar\'e plots,\cite{white12,vlad13,briguglio14,briguglio17}
which represent wave-particle phase shift (resonance detuning) and particle orbit excursion (radial decoupling) on the same footing.
In the kinetic Poincar\'e plot, each test particle's last completed orbit (when the particle crosses the equatorial plane at the outmost radial coordinate, i.e., poloidal angle $\theta=0$) is represented by a marker in the $(\Theta,P_{\phi})$ plane.
Here, $\Theta=\omega t_0-n\phi_0$ is the wave-particle phase at $\theta=0$, where $t_0$ and $\phi_0$ indicate the corresponding values of time and toroidal angle;
$P_{\phi}$ is the toroidal angular momentum given by Eq.~(\ref{eq:P_phi}).
Figure~\ref{fig:RSAE_HM}
shows the kinetic Poincar\'e plots for 3 successive time frames of the reference case, corresponding to, respectively, the linear stage, early nonlinear stage and saturation, as indicated in the energy evolution plot Fig.~\ref{fig:RSAE_HM}(d).
Here, note that $P_{\phi}$ is used to represent the test particle radial distribution, with larger $P_{\phi}$ corresponding to smaller $\bar r$ (orbit averaged radial coordinate) and vice versa.
In order to show more intuitively the mode width,
the radial mode structure is shown in Fig.~\ref{fig:RSAE_HM}(a) by mapping the test particle $\bar r$ into $P_{\phi}$ coordinates.
In the linear stage with negligible fluctuation amplitude, $P_{\phi}$ is conserved and the particles with $P_{\phi}=P_{\phi\rm res}$ (defined by $\omega_{\rm res}=\omega$) stay constant in phase.
Other particles with $P_{\phi}>P_{\phi\rm res}$ (red markers) and $P_{\phi}<P_{\phi\rm res}$ (blue markers) get a finite phase change after each bounce orbit, and thus, these markers drift along $\Theta$ in negative and positive direction, respectively.
When the fluctuation amplitude grows to a finite value, $P_{\phi}$ varies due to wave-particle interaction,
and an island-like structure naturally forms around the $P_{\phi\rm res}$, with increasing island width as the fluctuation amplitude grows.
The mode eventually saturates when the phase space structure extends over the region of effective linear wave-particle power transfer.
As clearly shown in Fig.~\ref{fig:RSAE_HM}(c), the resonant particles sample nearly the whole mode structure during their nonlinear orbit excursion,
suggesting that radial decoupling is a crucial element of the saturation mechanism.
As a quantitative assessment of the saturation mechanism,
Fig.~\ref{fig:RSAE_delta_r}
compares the averaged resonant test particle orbit radial excursion, $\Delta r_{\rm orbit}$, with $\Delta r_{\rm mode}$ and $\Delta r_{\rm power}$ for the cases reported in Fig.~\ref{fig:RSAE_nH}.
We can see that $\Delta r_{\rm orbit}$ is indeed similar to $\Delta r_{\rm power}$, and is comparable with $\Delta r_{\rm mode}$ for most of the cases.
Furthermore, as shown in Fig.~\ref{fig:RSAE_HM}, the mixing of particles from $P_{\phi}>P_{\phi\rm res}$ (smaller $\bar r$) with $P_{\phi}<P_{\phi\rm res}$ (larger $\bar r$) suggests a net outward particle flux due to the radial inhomogeneity of the EP distribution function.
The outward flux can also be shown from $C$ conservation since, as the particles lose energy to the wave, $P_{\phi}$ decreases ($r$ increases) and more particles move outward than inward during the mode growth stage.
Figure~\ref{fig:RSAE_density}
shows the distortion of test particles density profile at saturation for the reference case.
Consistent with the previous analysis, mode saturation is reached when the width of particle redistribution is comparable with the radial region of power transfer, since the resonant EP drive is significantly reduced.
Here, we emphasize that the fluctuation induced EP transport indeed occurs on meso-spatial scales [$\sim O(10^{-1}a)$], intermediate between macro-scales such as the equilibrium profiles [$\sim O(10^{0}a)$], and micro-scales of characteristic EP orbit width [$\sim O(10^{-2}a)$].
As anticipated in Ref.~\onlinecite{paper1}, the meso-scales reflect the properties of the fluctuation spectrum and the relevant toroidal mode number.\cite{zonca15a,chen16}
Note that the clear distortion of test particle density profile shown in Fig.~\ref{fig:RSAE_density} only reflects the considered $(M,C)$ slice of EP distribution function,
while the overall EP radial distribution, obtained by averaging the EP response over all the $(M,C)$ slices representing the entire phase space, is almost unchanged, due to the fact that most of the EPs are not resonant and, thus, experience a much weaker effect of the fluctuations.

\begin{figure}
\includegraphics{\myfigure 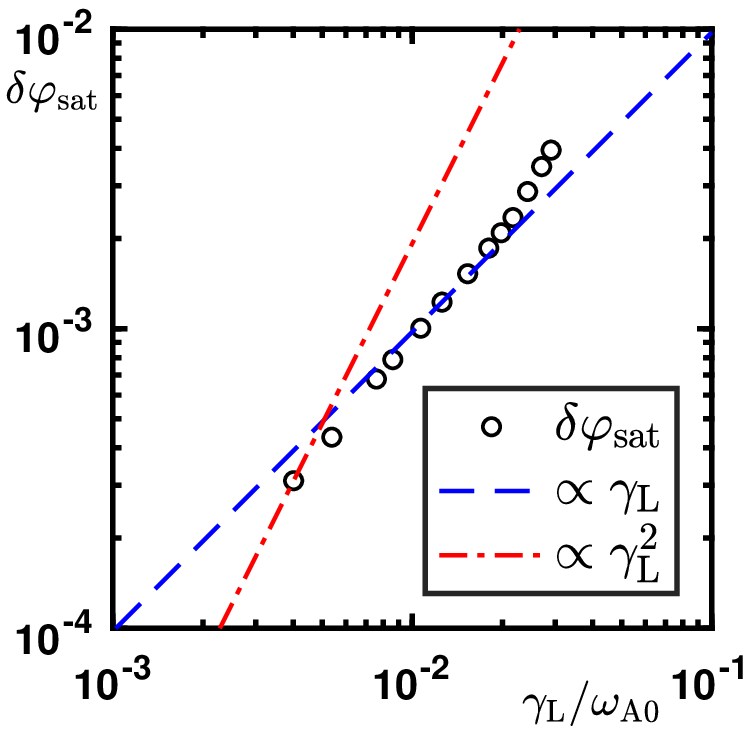}
\caption{For the cases reported in Fig.~\ref{fig:RSAE_nH}, the fluctuation amplitude (largest value in radial direction) at the initial saturation $\delta\varphi_{\rm sat}$, is shown as a function of the linear growth rate $\gamma_{\rm L}$ in logarithmic scale.
Here, $\delta\varphi_{\rm sat}$ is in units of $E_{0}/e_{\rm H}$ ($e_{\rm H}$ is the EP charge).
Dashed lines corresponding to linear and quadratic scaling are also indicated.}
\label{fig:RSAE_amplitude}
\end{figure}

\begin{figure}
\includegraphics{\myfigure 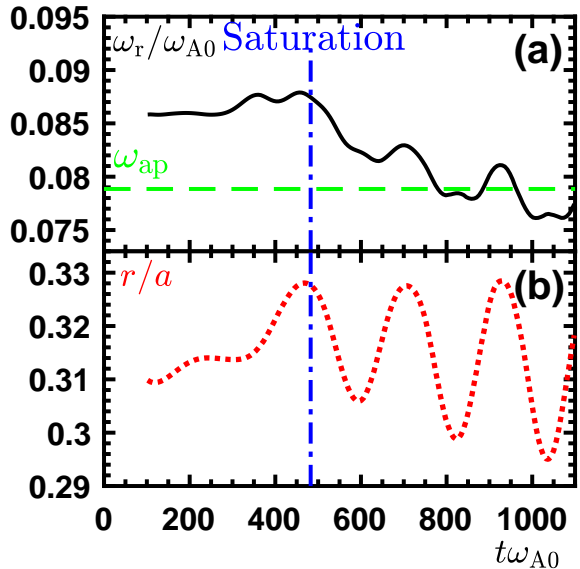}
\caption{Time evolutions of RSAE real frequency (maximum power intensity in the frequency spectrum) [frame (a)] and peak location of the radial mode structure [frame (b)] for a strongly unstable case with $n_{\rm H0}/n_{\rm i0}=3.0\times10^{-3}$.
The saturation time (vertical dash-dotted line) and the RSAE accumulation point frequency $\omega_{\rm ap}$ (horizontal dashed line) are also indicated.}
\label{fig:RSAE_freuquency}
\end{figure}

Figure~\ref{fig:RSAE_amplitude}
shows the saturated fluctuation amplitude versus $\gamma_{\rm L}$ for all considered cases.
We can observe that the scaling is clearly different from quadratic and is close to linear, as expected from radial decoupling being the dominant saturation mechanism.
In the low growth rate limit, where resonance detuning may also become important, we find a slight deviation towards a steeper scaling.
The approximately linear scaling is consistent with theoretical understanding\cite{zonca15b,chen16} and previous numerical simulations.\cite{wang16a,wang16b,briguglio17}
In our simulations, however, the scaling deviates from linear in the high growth rate limit.
This could be due to the fact that, in the strongly driven cases, the mode structures and frequencies are self-consistently modified
with the non-perturbative EP redistribution.\cite{zonca15a,zonca15b,chen16}
As an example, Fig.~\ref{fig:RSAE_freuquency}
shows the time evolutions of mode frequency and fluctuation radial peak location (representative of the mode structure) of a strongly unstable case.
The self-consistent modulation of mode frequency and mode structure becomes evident when the fluctuation grows to an appreciable amplitude approaching saturation.
Simulation results suggest that fluctuations are further enhanced in this self-consistent non-perturbative process, and saturate at a higher amplitude than the predicted linear scaling, which assumes constant mode frequency and frequency-independent mode structure.\cite{wang16a,wang16b,briguglio17}
From Fig.~\ref{fig:RSAE_freuquency}, it is also interesting to note that, on longer time scale of the strongly unstable case with clear non-perturbative wave-EP interactions,
the frequency chirping shows non-adiabatic features, as the mode structure is strongly modified.
In such conditions, it is expected that the non-perturbative EP redistributions may become secular and characterized by avalanches,\cite{white83,zonca05} which are important issues in burning plasma physics studies.\cite{zonca15a,zonca15b,chen16}
Thus, a more detailed analysis of these behaviors is worthwhile being pursued and will be continued in future work, since it is beyond the intended scope of this paper, which mainly aims at illuminating the diverse and rich nonlinear physics that can be investigated in DTT.
Here, we just note that the nonlinear dynamics of strongly unstable cases further addresses the importance of self-consistent treatment of mode structure and EP nonlinear evolutions, especially in the next generation tokamak relevant conditions.

\subsection{TAE nonlinear dynamics}
\label{subsec:outer}

\begin{figure}
\includegraphics{\myfigure 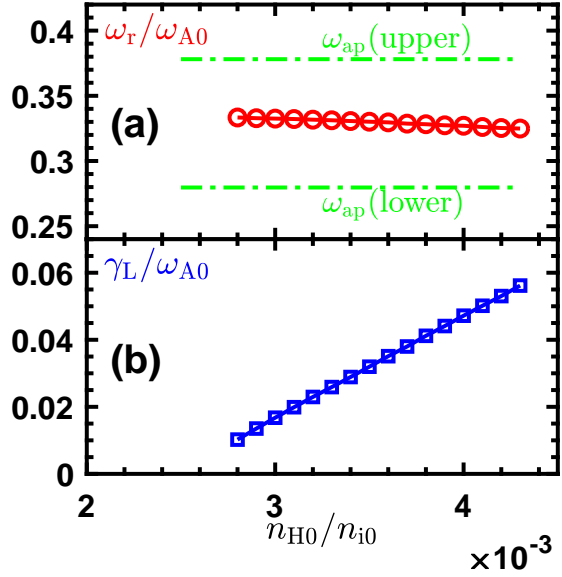}
\caption{TAE real frequencies $\omega_{\rm r}$ [frame (a)] and linear growth rates $\gamma_{\rm L}$ [frame (b)] shown as functions of EP on-axis density $n_{\rm H0}$.
In addition, the upper and lower TAE accumulation point frequencies $\omega_{\rm ap}$ (closest to mode peak) are indicated as horizontal dash-dotted lines to illustrate the importance of non-local coupling with SAW continuum.}
\label{fig:TAE_nH}
\end{figure}

\begin{figure}
\includegraphics{\myfigure 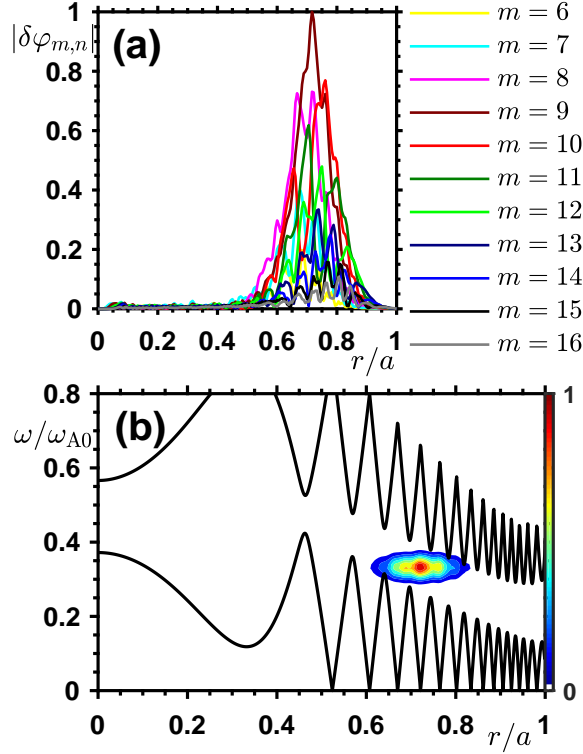}
\caption{Radial TAE mode structure for several Fourier decomposed poloidal harmonics of scalar potential fluctuation $\delta\varphi_{m,n}$ [frame (a)], and the intensity contour plot of the corresponding power spectrum in the $(r,\omega)$ plane [frame (b)] for the ``weak TAE case'' with $n_{\rm H0}/n_{\rm i0}=3.0\times10^{-3}$ in the linear stage. The solid curves in frame (b) represent SAW continua.}
\label{fig:TAE_mode}
\end{figure}

\begin{figure*}
\includegraphics{\myfigure 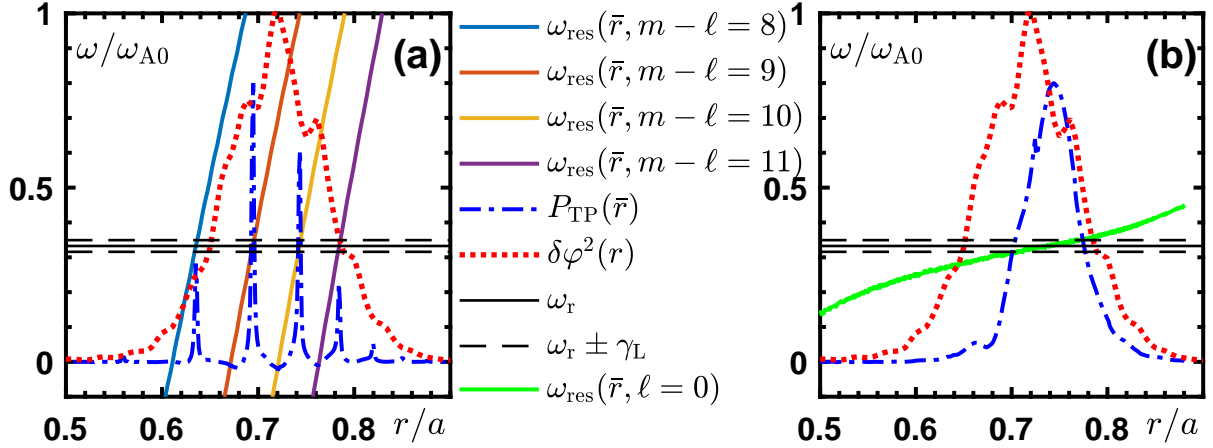}
\caption{Linear structures of several transit harmonic resonances by co-passing particles [frame (a)] and the precession resonance by trapped particles [frame (b)] for the weak TAE case.
Analogous to Fig.~\ref{fig:RSAE_resonance}, the radial profiles of test particle power transfer $P_{\rm TP}(\bar r)$ in each analysis and the radial mode structure $\delta\varphi^2(r)$ are also indicated.
Note that, in frame (a), the transit resonances are identified by the effective resonance harmonic $m-\ell$.}
\label{fig:TAE_resonance}
\end{figure*}

Contrary to the weakly damped RSAE fluctuations analyzed above, the TAE fluctuations in the outer core region experience heavy damping due to strong coupling with the SAW continuum.\cite{zonca92,rosenbluth92,zonca93}
Therefore, larger values of EP density are applied to drive the TAE fluctuations unstable.
Figure~\ref{fig:TAE_nH}
shows $\omega_{\rm r}$ and $\gamma_{\rm L}$ of the $n=6$ TAE fluctuations with $n_{\rm H0}/n_{\rm i0}$ in the range of $[0.0028,~0.0043]$,
where the relatively high destabilization threshold (with respect to the RSAE fluctuations in the central core region) can be clearly observed from the trend of $\gamma_{\rm L}$.
First, we focus on a weakly unstable case with $n_{\rm H0}/n_{\rm i0}=3.0\times10^{-3}$ (we will refer to this case as ``weak TAE case'' in the following), whose mode structure in the linear stage is shown in Fig.~\ref{fig:TAE_mode}.
The mode structure appears as a broad radial envelope consisting of a wide range of coupled poloidal harmonics, as expected from the equilibrium profiles and the structure of SAW continua.
As discussed in Ref.~\onlinecite{paper1} and introduced above, the peculiar interest for the TAE fluctuations is that different types of EPs in the velocity space distribution provide finite mode drive via their respective resonances: precession/transit resonances for trapped/passing EPs.
On account of the fact that the TAE fluctuations are typically localized in the outer core region, they may be more of a concern for EP confinement, due to their potential effect of causing significant outward EP flux and consequently, of damaging the plasma facing components.
Furthermore, the radial structure of wave-passing EP power transfer appears as several isolated peaks, as shown in Fig.~\ref{fig:TAE_resonance}(a)
by test particle analysis.
Here, for simplicity, only co-passing EPs are analyzed in detail, as the behaviors of co- and counter-passing EPs are very similar due to similar resonance structures.
Note that in Fig.~\ref{fig:TAE_resonance}(a), the transit resonances are identified by the effective resonance harmonic $m-\ell$ from the particle's perspective, since both the poloidal harmonic $m$ and ``bounce'' harmonic $\ell$ enter in the expression of transit resonance frequency as the combination $m\pm\ell$ [cf. Eq.~(\ref{eq:omega_res_passing})], with minus/plus sign for co-/counter-passing particle, respectively.
However, at each resonant radius characterized by $m-\ell$ in Fig.~\ref{fig:TAE_resonance}(a), multiple poloidal harmonics of the mode are excited via the corresponding transit harmonic resonances,
which are weighted differently by finite orbit width bounce averaging,
consistent with the mode structure shown in Fig.~\ref{fig:TAE_mode}(a).
Thus, toroidal mode number and finite normalized (with respect to plasma minor radius) EP orbit width play crucial roles.
In addition, we note that the characteristic scale length of the radial separation of the transit harmonic resonances is $1/nq'$; that is, the meso-scale of drift wave turbulence, typically characterized by much higher toroidal mode number.\cite{zonca14c,zonca15a,zonca15b,chen16,briguglio17,paper1}
On the other hand, the resonance structure of trapped EPs, shown in Fig.~\ref{fig:TAE_resonance}(b), is similar to the RSAE case analyzed in Sec.~\ref{subsec:central}.
We can see that $\Delta r_{\rm power}$ of passing particles is limited by the resonance condition, and is much narrower than $\Delta r_{\rm mode}$.
The situation is not as clear for trapped particles, however, it is still legitimate to expect that resonance detuning, more than radial decoupling, is a relevant saturation mechanism for the weak TAE case.

\begin{figure}
\includegraphics{\myfigure 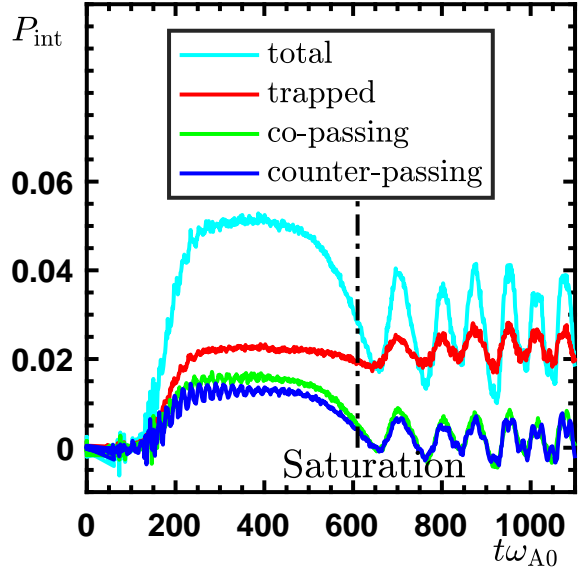}
\caption{Time evolution of phase space integrated power transfer for the weak TAE case.
The power transfer is normalized to the sum of instantaneous kinetic and magnetic energy density.
The saturation time is indicated as a vertical dash-dotted line.}
\label{fig:TAE_power1}
\end{figure}

\begin{figure*}
\includegraphics{\myfigure 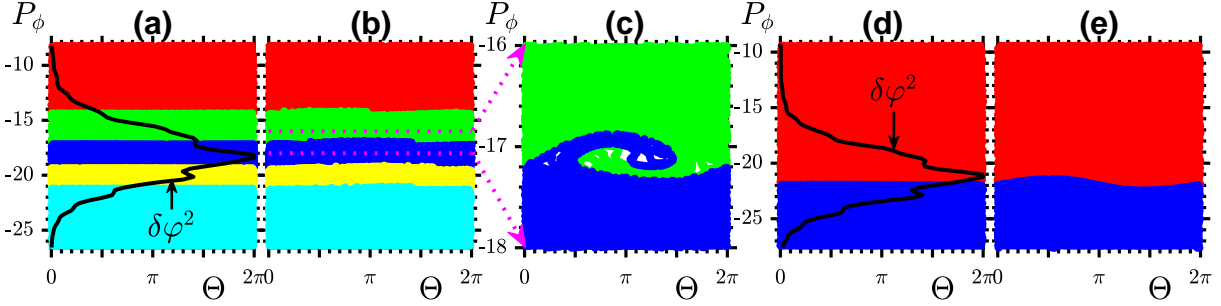}
\caption{Kinetic Poincar\'e plots of co-passing [frames (a)-(c)] and trapped test particles [frames (d)-(e)] for the weak TAE case, where frames (a) and (d) refer to the linear stage and frames (b) and (e) to the saturation time (see Fig.~\ref{fig:TAE_power1}).
As labeled on the figure, frame (c) is a zoom of frame (b) to more clearly visualize the region affected by one single resonance therein.
Analogous to Fig.~\ref{fig:RSAE_HM}, the test particle markers are colored by their initial values of $P_\phi$, with different colors denoting the transition across the resonances shown in Fig.~\ref{fig:TAE_resonance}.
In addition, radial mode structure is shown in frames (a) and (d) by mapping $\bar r$ into $P_\phi$ coordinates.}
\label{fig:TAE_HM1}
\end{figure*}

Due to the different resonance structures of trapped and passing EPs and, in particular, the resonant interaction length scales, they play different roles in the nonlinear saturation of TAE fluctuations.
Figure~\ref{fig:TAE_power1}
shows the time evolution of phase space integrated power transfer for the weak TAE case.
We can observe that, mode saturation is due to significant reduction of passing particle drive.
Meanwhile, trapped particle drive is still kept in a significant level, suggesting that the strong damping also plays an important role in mode saturation, as the residual trapped particle drive approximately balances the dissipation.
Thus, the passing particle resonance plays a more crucial role in mode saturation for the weak TAE case.
Figure~\ref{fig:TAE_HM1}
shows the kinetic Poincar\'e plots of both passing and trapped particles in the linear stage and at saturation, with different colors denoting the transition across the resonances shown in Fig.~\ref{fig:TAE_resonance}.
We can observe that the mode saturates when the resonant passing EP orbit excursion matches the narrow radial width of linear power transfer.
In fact, since $\Delta r_{\rm power}{\rm(passing)}$ is very small, the phase space structure of passing particles can be clearly seen on the zoomed scale in Fig.~\ref{fig:TAE_HM1}(c) only, which is shown as an expanded insert of the phase space region affected by one single resonance in Fig.~\ref{fig:TAE_HM1}(b).
Moreover, the redistribution of resonant passing particles is very weak and localized around $P_{\phi\rm res}$;
no interaction of adjacent resonances takes place, since the perturbation of equilibrium particle orbits is exceedingly small.
(Note that, this is the criterion of ``weak drive''.)
On the other hand, trapped particle nonlinear transport is intrinsically nonlocal [cf. Fig.~\ref{fig:TAE_resonance}(b)].\cite{zonca15b,chen16}
However, as a result of the low fluctuation amplitude, trapped particle transport also occurs on a much smaller scale compared to the mode width, similar to the case in Fig.~\ref{fig:RSAE_HM}(b).
The fluctuation induced particle transport, thus, does not have significant impact on the power transfer by trapped EPs.
The relative ordering of resonant particle response length scale and the mode width suggests that resonance detuning is indeed the dominant saturation mechanism for the weak TAE case, in contrast to the RSAE cases with similar growth rate that is regulated by radial decoupling mechanism.
Therefore, our simulation results suggest that the underlying mechanism of mode saturation and EP nonlinear dynamics is not only determined by the linear growth rate but, more crucially, by the wave-EP resonance structures, where the saturation mechanism is reflected by the relative ordering of the scale lengths of mode structure, wave-EP resonant power transfer and nonlinear EP transport.\cite{briguglio14,zonca15a,chen16}
Furthermore, the clear diversity of RSAE fluctuations in the central core region and TAE fluctuations in the outer core region also illustrates the capability of DTT to address a variety of nonlinear EP physics related with burning plasma studies.

\begin{figure}
\includegraphics{\myfigure 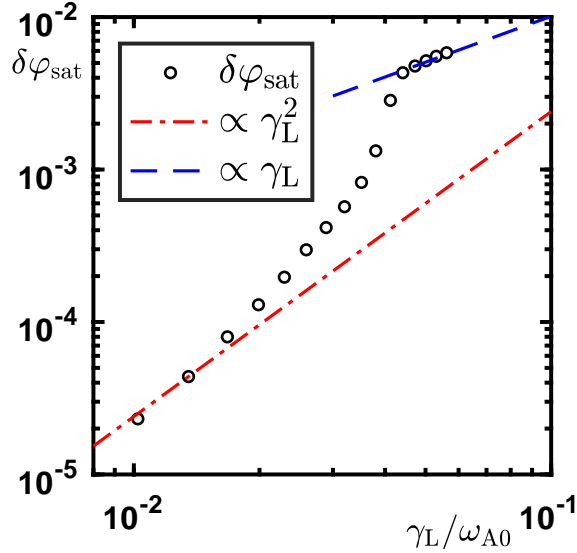}
\caption{For the cases reported in Fig.~\ref{fig:TAE_nH}, the saturated fluctuation amplitude $\delta\varphi$ is shown with respect to $\gamma_{\rm L}$ in logarithmic scale.
Quadratic and linear scalings in different regimes are indicated by dashed lines.}
\label{fig:TAE_amplitude}
\end{figure}

\begin{figure}
\includegraphics{\myfigure 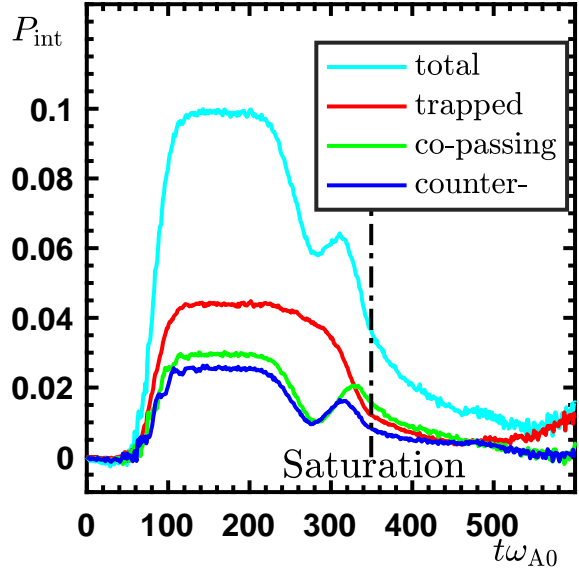}
\caption{Time evolution of and integrated power transfer for the ``strong TAE case''.
The saturation time is indicated as a vertical dash-dotted line.}
\label{fig:TAE_power2}
\end{figure}

Figure~\ref{fig:TAE_amplitude}
shows the saturated fluctuation amplitude with respect to $\gamma_{\rm L}$ for all TAE cases reported in Fig.~\ref{fig:TAE_nH}.
In the low growth rate limit, the scaling is close to quadratic, confirming that resonance detuning is indeed the main saturation mechanism.
Moreover, in the higher growth rate cases, the scaling first becomes higher than quadratic, and reduces to approximately linear in the strongly unstable limit.
It suggests that qualitative and quantitative differences of wave-EP nonlinear dynamics take place with increasing EP drive, due to equilibrium geometry, plasma nonuniformity and non-perturbative EP response.\cite{briguglio98,wang12,zonca15b,chen16}
As an example of the strongly unstable regime,
we look at the case with $n_{\rm H0}/n_{\rm i0}=4.0\times10^{-3}$ (in the following, we refer to this case as ``strong TAE case'').
The strong TAE case shows similar linear mode and resonance structures compared to those of the weakly unstable case shown in Figs.~\ref{fig:TAE_mode} and \ref{fig:TAE_resonance}.
The power transfer widths of both passing and trapped particles are larger in the strong TAE case as $\Delta r_{\rm power}{\rm (passing)}\ll\Delta r_{\rm power}{\rm (trapped)}\simeq\Delta r_{\rm mode}$, due to increased $\gamma_{\rm L}$.
(Note that the several transit harmonic resonances are still well separated in the linear stage.)
However, as shown in Fig.~\ref{fig:TAE_power2},
the mode time evolution has different features.
We note that, after the linear growth stage, during $t\sim240-280\omega_{\rm A0}^{-1}$, the total power transfer decreases, mostly due to a rapid reduction of passing EP drive, since the trapped EP drive is not significantly impacted in this stage.
Different from the weak TAE case, the mode keeps growing due to less affected trapped EP drive and residual passing EP contribution.
Then, a short second growth stage follows by an interestingly strengthened passing EP drive; meanwhile, trapped EP drive starts clearly decreasing.
At $t\sim350\omega_{\rm A0}^{-1}$, the mode eventually saturates with significant decrease of both passing and trapped EP drive.

\begin{figure*}
\includegraphics{\myfigure 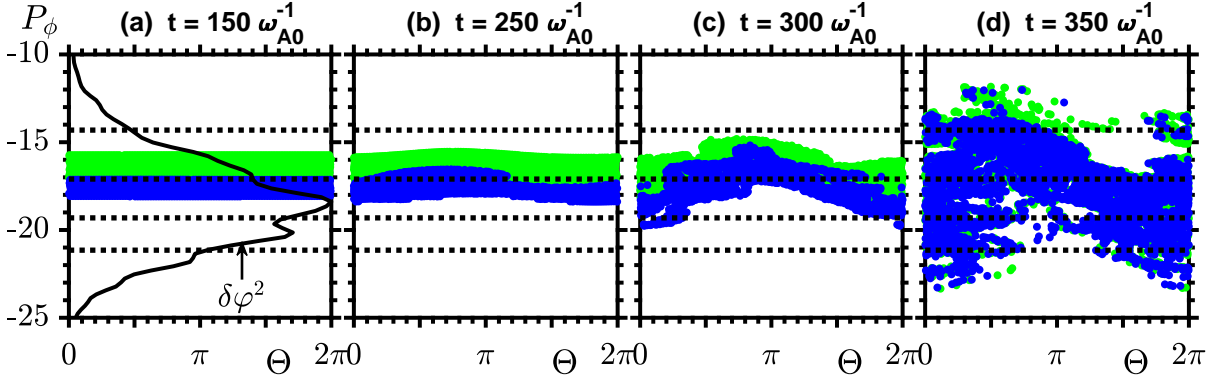}
\caption{Kinetic Poincar\'e plots of co-passing test particles at four times of the strong TAE case with test particles initialized around the $m-\ell=9$ resonance.
The four times refer to the linear stage [frame (a)], two times of the nonlinear growth stage [frames (b) and (c)] and saturation time [frame (d)].
The radial mode structures is shown in frame (a) by mapping $\bar r$ into $P_{\phi}$ coordinates.
In all frames, linear $P_{\phi\rm res}$ with $m-\ell=8\div11$ are indicated by horizontal dotted lines.}
\label{fig:TAE_HM2}
\end{figure*}

More details underlying this complicated time evolution can be illustrated by test particle analysis, where we focus on the novel nonlinear dynamics of passing EPs.
Figure~\ref{fig:TAE_HM2}
shows kinetic Poincar\'e plots of co-passing test particles at four times of the strong TAE case.
Here, in order to see the nonlinear dynamics more clearly, the test particles are distributed around one single resonance with $m-\ell=9$ at $P_{\phi\rm res}\simeq-17.10 m_{\rm H}av_{\rm H}$.
In addition, linear $P_{\phi\rm res}$ with $m-\ell=8\div11$ are also indicated in $P_\phi$ coordinate.
The dynamics in the early nonlinear stage [Fig.~\ref{fig:TAE_HM2}(b)] is similar to the weak TAE case analyzed above, that is, the passing particles are radially redistributed around $P_{\phi\rm res}$, and the power transfer by passing particles decreases correspondingly.
However, since the fluctuation strength keeps increasing mostly due to magnetically trapped EP drive, the resonant island extends and passing particle transport becomes nonlocal, as the particles are distributed to an increasingly wider region.
At $t\sim300\omega_{\rm A0}^{-1}$ [Fig.~\ref{fig:TAE_HM2}(c)],
we can observe that the particle distribution is strongly distorted,
where a substantial part of particles are radially transported on a radial scale comparable with separation of adjacent resonances ($1/nq'$).
Thus, the expanding resonant islands, whose characteristic widths scale as $\sqrt{\delta\varphi}$, are effectively overlapping.
Since the wave-passing EP resonant interaction scale length is very narrow in the linear stage,
the meso-scale transport and increasing resonant island width allow the wave to more effectively extract energy from particles, including the ones that are not resonant in the linear stage, as they still retain a significant amount of free energy.\cite{zonca15b}
In fact, all particles with significantly modified phase space orbits [cf. Fig.~\ref{fig:TAE_HM2}(d)] are resonant in this stage, and their small but finite resonant drive contributes to the increase in the integrated power transfer of passing particles shown in Fig.~\ref{fig:TAE_power2}.
This enhanced mode drive is also responsible for the stronger (than quadratic) scaling of saturation amplitude with $\gamma_L$ in Fig.~\ref{fig:TAE_amplitude}.\cite{briguglio98,wang12,chen16}
Eventually at mode saturation [Fig.~\ref{fig:TAE_HM2}(d)], we notice that the radially redistributed particles almost sample the whole mode structure, with the corresponding decrease of power transfer.

\begin{figure}
\includegraphics{\myfigure 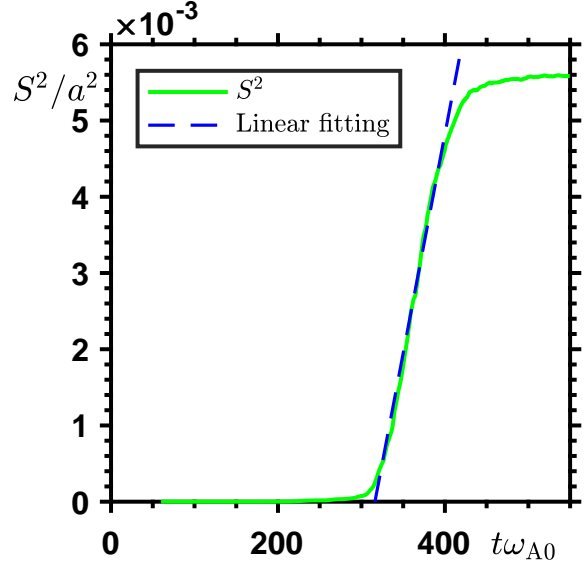}
\caption{Time evolution of the mean squared displacement $S^2$ for
a group of resonant passing particles initialized with $P_{\phi}=P_{\phi\rm res}\simeq-17.10m_{\rm H}av_{\rm H}$ and evenly distributed on the flux surface.}
\label{fig:TAE_msd}
\end{figure}

It is also interesting to further address the mechanism of resonant passing particle transport at mode saturation.
As shown in Fig.~\ref{fig:TAE_HM2}(d), the particle transport is significantly enhanced in this stage due to the high fluctuation amplitude, and it is difficult to identify any corresponding phase space structure.
This happens because the particles are transported on a spatial scale larger than or comparable with the radial resonance separation $1/nq'$, with the consequent overlap of adjacent resonances (Chirikov condition).
Thus, as particles receive random ``kicks'' from overlapping resonances, their nonlinear orbits become stochastic in this stage.
As a result of orbit stochasticity, the nature of passing particle transport becomes diffusive.
In order to illustrate the transition from localized redistribution to meso-scale diffusive transport more clearly,
Fig.~\ref{fig:TAE_msd}
shows the time evolution of the mean squared displacement $S^2$ for
a group of resonant passing test particles initialized with $P_{\phi}=P_{\phi\rm res}$ and evenly distributed on the flux surface.
Here, test particle's outer equatorial plane radial coordinate $r_0$ is used to represent the particle's radial position, and $S^{2}$ is calculated as
\[
S^{2}(t)=\left<\left(r_0(t)-\langle r_0(t)\rangle\right)^2\right>,
\]
where $\langle\dots\rangle$ stands for averaging over the test particle population.
Thus, $S^{2}$ is the characteristic relative radial separation of the test particles, and is representative of particle transport length scales.
In the linear stage, $S^{2}\simeq0$ since all particles have the same $r_0$.
In the early nonlinear stage, the particles are re-distributed around $P_{\phi\rm res}$, and the collective particle transport remains coherent with relatively low value of $S^2$.
When fluctuation amplitude reaches a threshold value, and the scale of particle transport increases to the resonance separation length scale,
we notice that $S^2$ increases significantly.
When this occurs, $S^2$ scales roughly linearly with time, suggesting that the nature of particle transport is indeed diffusive as anticipated above.
The diffusion rate $D$ can be estimated by the slope of $S^2(t)$ as
\[
D\simeq\frac{1}{2}\frac{{\rm d} S^2}{{\rm d} t}.
\]
After mode saturation, $S^2$ also reaches a steady state value characterized by the mode envelope width, as the particles are roughly evenly distributed within the mode location:
$S_{\rm sat}^{2}\sim1/12(\Delta r_{\rm mode})^2\simeq5.2\times10^{-3}a^2$.
Thus, the finite mode width becomes effective in preventing the particles from being transported further out.

\begin{figure}
\includegraphics{\myfigure 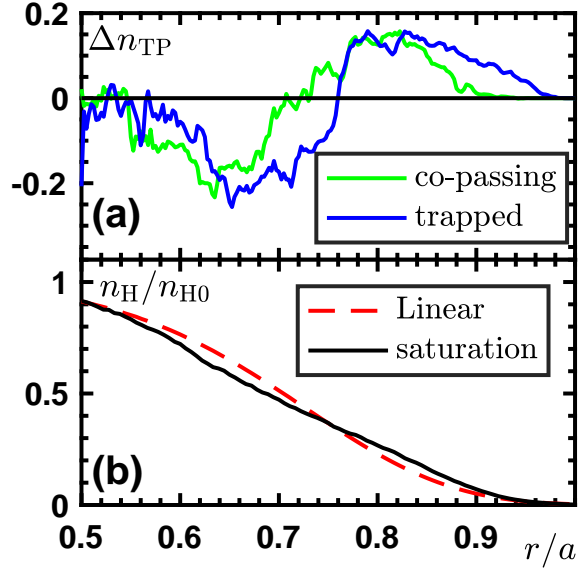}
\caption{The distortion of test particle density profile $\Delta n_{\rm TP}$ [frame (a)] shown in arbitrary unit for the strong TAE case.
The calculation and normalization of $\Delta n_{\rm TP}$ are analogous to Fig.~\ref{fig:RSAE_density}.
Frame (b) shows the comparison of integrated EP density $n_{\rm H}$ in the linear stage (dashed curve) and at saturation (solid curve).}
\label{fig:TAE_density}
\end{figure}

The meso-scale EP transport and high saturation amplitude also result in significant EP redistribution in the strong TAE case.
Figure~\ref{fig:TAE_density}(a)
shows the distortion of test particle density profile at saturation for both co-passing and trapped particles.
We can see that at mode saturation, the outward particle fluxes of both types of particles occur on similar scales comparable with the mode width, consistent with the significant reduction of power transfer in both channels.
Since a substantial portion of the EPs are resonant, the integrated EP density $n_{\rm H}$ also exhibits a clear outward EP flux at saturation, as shown in Fig.~\ref{fig:TAE_density}(b).
Thus, the TAE fluctuation induced EP loss could be more crucial for the assessment of EP confinement, and such a problem is worthwhile being further analyzed in future works, taking into account the whole Alfv\'en fluctuation spectrum self-consistently.
Furthermore, since the finite mode width is the more effective factor in regulating the nonlinear transport of both types of particles, radial decoupling is the dominant saturation mechanism for the strong TAE case, confirmed also by the approximately linear scaling of saturation amplitude in the high growth limit of Fig.~\ref{fig:TAE_amplitude}.

\section{Summary and discussion}
\label{sec:sumdis}

In this paper, following the previous work in Ref.~\onlinecite{paper1}, we have analyzed the nonlinear dynamics of shear Alfv\'en fluctuation saturation and the corresponding fluctuation induced EP transport in DTT plasmas.
The simulations address two particular cases, namely, $n=4$ RSAE fluctuations and $n=6$ TAE fluctuations for the central and outer core regions of DTT plasmas, respectively.
These cases can be considered as typical paradigms to illustrate the rich and diverse physics due to resonant excitation of Alfv\'enic fluctuations by supra-thermal particles in DTT and, more generally, in reactor relevant fusion plasmas.
In particular, we focus on the mode saturation mechanism and on the relative importance of resonance detuning versus radial decoupling, exploring the properties of EP transport as coherent redistribution and/or diffusive transport.
Test particle method and Hamiltonian mapping technique are extensively used to illustrate the wave-EP resonant interactions and the nonlinear evolution of EP phase space zonal structures.

The nonlinear saturation of RSAE fluctuations in the central core region, dominated by trapped particle precession resonance, is consistent with previous theoretical and numerical studies.
By analyzing the linear mode and resonance structures, nonlinear EP orbit excursion as well as saturation amplitude, we show that the prevalent saturation mechanism is radial decoupling,
which plays an important role even when the fluctuation is close to marginal instability.
This is consistent with theoretical understanding, where the effect of resonance detuning is expected to be much weaker for trapped than passing particles, due to  the flatter radial profile of the resonance frequency.
Moreover, the radial scale of resonant EP re-distribution is generally comparable with the mode radial width.
Thus, the fluctuation induced trapped EP transport expectedly occurs on meso-spatial scales for a wide range of reference equilibria and corresponding plasma stability.
In fact, the relevant spatio-temporal scales of nonlinear wave-EP dynamics are controlled by equilibrium geometry, plasma nonuniformity and perturbative versus non-perturbative EP response; and are ultimately reflected by the features of the fluctuation spectrum and the corresponding most unstable toroidal mode number.
The scaling of saturated fluctuation amplitude with respect to the linear growth rate is, in general, approximately linear.
Meanwhile, in high growth rate limit, self-consistent modulations of mode structure and frequency are observed, which maximize wave-EP power transfer and contribute to the enhanced saturation level in this regime.
In addition, the longer time scale evolution with strong drive suggests non-adiabatic frequency sweeping and secular EP transport, similar to those of strongly unstable EPMs.
A detailed investigation of longer time scale nonlinear dynamics is beyond the scope of this work, and will be the subject of a future publication.
In fact, we note that the strongly unstable RSAEs discussed in this work are in the relevant parameter regime for burning plasma physics studies.\cite{paper1,pizzuto10}

The nonlinear dynamics of TAE fluctuations in the outer core region can be quite different depending on the strength of mode drive.
The TAE fluctuations are driven collectively by both magnetically trapped and passing EPs, which show very different resonant interaction length scales due to their resonance condition: wave-particle power transfer with passing EPs is characterized by much finer scale than trapped EPs.
Thus, trapped and passing EPs exhibit different nonlinear transport length scales and play independent or synergetic roles in mode saturation under various stability regimes.
For sufficiently low linear growth rate, low amplitude saturation is observed, mainly due to the nonlinear reduction of passing EP drive, while trapped EP drive is essentially unaltered.
EP radial redistribution in this case is local in phase space; thus, transit resonances are radially well separated and transport effects on the EP density profile are negligible.
Meanwhile, for stronger linear growth rate, with sufficiently high fluctuation level and corresponding enhanced EP radial excursion, transit resonances may overlap.
Phase space orbits become stochastic and passing EP radial transport is diffusive over the length scale of the mode width.
The fluctuation saturates with meso-scale redistributions over the whole radial mode structure for both resonant trapped and passing EPs;
and the overall EP flux is reflected by a significant distortion of the EP density profile.
This suggests that TAEs in the outer core region may be a more serious concern than RSAEs in the central core for the limits they may impose on plasma operations to avoid global EP losses.
Furthermore, note that for single-$n$ simulations reported in this paper, large power input and the contribution of trapped EP drive are necessary to cause transit resonance overlap and diffusive transport of passing EPs.
For realistic scenarios with multi-$n$ modes excited simultaneously, the stochasticity threshold is much lower with much more resonances.\cite{feng13}
Thus, EP diffusive transport by spontaneously excited multi-$n$ TAEs could occur at much lower EP concentration, and will be further explored in future studies.
The present work also suggests that transition to resonance overlap and diffusive EP transport is connected with equilibrium geometry and plasma nonuniformity as well as non-perturbative EP response.
In fact, stronger EP drive causes the saturation level to be enhanced over the quadratic scaling with the linear growth rate to be expected for resonance detuning.
The scaling finally reduces to approximately linear in the high growth rate limit with nonlinear EP transport comparable to the mode width,
suggesting that radial decoupling should be expected for strongly driven TAEs in the outer core region.

In summary, by further investigating the DTT reference scenario assumed in Ref.~\onlinecite{paper1}, the present work confirms the anticipations on the rich and diverse physics that is expected in DTT core plasmas.
The characterizing element is the Larmor radius normalized to plasma minor radius, $\rho^*\equiv\rho_{\rm L}/a$, for both EPs as well as thermal plasmas.
In particular, the ratio of these two fundamental parameters, which is controlled by the characteristic EP energy in units of the critical energy, plays a fundamental role.
In DTT plasmas and, more generally, in reactor relevant conditions, the micro-scales of Alfv\'enic instabilities resonantly excited by EPs are of the same order of meso-scale structures due to drift wave turbulence.
This is one crucial reason why EPs are considered mediators of cross scale couplings, with their predominant contribution to the local power balance further emphasizing their unique role.

Another important physics process illuminated by the present work is the nonlocal transfer of energy and momentum in phase space, due to the peculiar role of magnetically trapped and passing EPs. This {\em channeling} in phase-space, which may involve different mode numbers, and the general properties of the fluctuation spectrum discussed in this work, confirm the importance of looking at transport processes in phase space when dealing with collisionless fusion plasmas; that is, the importance of phase space zonal structures.

\acknowledgments

The authors would like to thank G. Fogaccia for precious support in the simulation works.
This work was supported by National Key R\&D Program of China under Grant No. 2017YFE0301900.
This work was also carried out within the framework of the EUROfusion Consortium and received funding from the EURATOM research and training programme 2015–2018 under Grant Agreement No. 633053 (Project Nos. WP15- ER/ENEA-03 and WP17-ER/MPG-01).
The views and opinions expressed herein do not necessarily reflect those of the European Commission.
The computing resources and the related technical support used for this work have been provided by CRESCO/ENEAGRID High Performance Computing infrastructure and its staff.\cite{CRESCO}

\appendix

\section{Test particle selection}
\label{app:TPs}

Following Ref.~\onlinecite{briguglio14}, the test particle selection is introduced in this appendix.
In the equilibrium magnetic field, the particle toroidal angular momentum $P_{\phi}$ and energy $E$ are constants of motion.
However, in the presence of a finite amplitude fluctuation, due to wave-particle interactions (e.g., ${\bf E}\times{\bf B}$ drift), the particle orbit is perturbed and conservations of $P_{\phi}$ and $E$ are broken.
Following Refs.~\onlinecite{white12,zonca15b,chen16},
for a single-$n$ mode with constant frequency, a new quantity can be constructed from the extended phase space Hamiltonian,
\begin{equation}
C\equiv\omega P_{\phi}-nE,
\label{eq:C}
\end{equation}
which is conserved in addition to the magnetic moment $M$.
In the physics model of HMGC,
\begin{equation}
P_{\phi}\simeq m_{\rm H}RU+e_{\rm H}R_{0}(\psi_{\rm eq}-\psi_{\rm eq0})/c
\label{eq:P_phi}
\end{equation}
at the leading order.\cite{briguglio14}
Here, $R$ is the major radius coordinate,
$U$ is the parallel (to the equilibrium magnetic field) velocity,
$e_{\rm H}$ is the EP charge,
$\psi_{\rm eq}$ is the equilibrium magnetic flux function,
and $\psi_{\rm eq0}$ is the value of $\psi_{\rm eq}$ on the magnetic axis.
Moreover, $E=m_{\rm H}U^{2}/2+M\Omega_{\rm H}$.

Given the conservation properties of $M$ and $C$, we could then only look at a single resonant ``slice'' ($M_{0}, C_{0}$) of the EP distribution function,
since linear and nonlinear evolution of the considered slice is independent of others.
The selected slice is identified from a reduced phase space grid $(r,M,U)$ with significant wave-EP power transfer in the linear stage,
and is sampled by a group of test particles, which are initialized with $M_{0}$, $C_{0}$ and properly (e.g., uniformly) distributed in $r$, $\theta$, $\phi$ directions,
where $\theta$ and $\phi$ are respectively, poloidal and toroidal angles.
Test particles are evolved in the electromagnetic field stored from the self-consistent simulation and, thus, are representative of the dynamic behavior of physical particles with the same phase space coordinates.
The test particle characteristic resonance frequency, $\omega_{\rm res}$, can be computed as\cite{zonca15b}
\begin{equation}
\omega_{\rm res}(\bar r,M_{0},C_{0},\ell)=n\omega_{\rm d}+\ell\omega_{\rm b}
\label{eq:omega_res_trapped}
\end{equation}
for magnetically trapped particles, and as
\begin{equation}
\omega_{\rm res}(\bar r,M_{0},C_{0},\ell)=n\omega_{\rm d}+\ell\omega_{\rm b}+(n\bar q-m)\sigma\omega_{\rm b}
\label{eq:omega_res_passing}
\end{equation}
for passing particles.
Here, $\bar r$ stands for the orbit averaged particle radial coordinate (playing the same role as $P_\phi$ for fixed $M_0$ and $C_0$);
$\omega_{\rm d}=(\Delta\phi/2\pi-\sigma\bar q)\omega_{\rm b}$ is the toroidal precession frequency, where $\Delta\phi$ in the change of $\phi$ over the period of a bounce orbit, $\tau_{\rm b}=\oint{\rm d}\theta/\dot\theta$, $\sigma={\rm sgn}(U)$, $\bar q$ is the weighted safety factor integrated along the particle orbit;\cite{zonca15b}
$\ell$ is the bounce harmonic;
and $\omega_{\rm b}=2\pi/\tau_{\rm b}$ is bounce/transit frequency for trapped/passing particles.
Furthermore, for a mode with finite linear growth rate $\gamma_{\rm L}$, the condition for effective resonant power transfer could be given as
\begin{equation}
|\omega-\omega_{\rm res}(r,M_{0},C_{0},\ell)|\lesssim\gamma_{\rm L}.
\label{eq:resonance}
\end{equation}
That is, significant wave-particle resonant interaction can take place when the frequency difference is of the order of $\gamma_{\rm L}$.
Eq.~(\ref{eq:resonance}) could be solved with respect to $r$ or $P_{\rm \phi}$, yielding the resonance width $\Delta r_{\rm res}(\omega,M_{0},C_{0},\ell)$ or $\Delta P_{\rm \phi res}(\omega,M_{0},C_{0},\ell)$.

%

\end{document}